\documentclass[preprint, review, 12pt]{elsarticle}

\usepackage{amssymb}
\usepackage{fullpage}
\usepackage{amsthm}
\usepackage{amsmath}
\usepackage{caption}
\usepackage{subcaption}
\usepackage{float}
\usepackage{hyperref}
\usepackage{bm}
\usepackage{xcolor}

\begin{document}

\begin{frontmatter}

\title{Simulation of non-Gaussian wind field as a $3^{rd}$-order stochastic wave}

\author{Lohit Vandanapu}
\author{Michael D. Shields}

\address{Johns Hopkins University, Baltimore, United States}

\begin{abstract}
This paper presents a methodology for the simulation of non-Gaussian wind field as a stochastic wave using the 3rd-order Spectral Representation Method.  Traditionally, the wind field is modeled as a stochastic vector process at discrete locations in space. But the simulation of vector process is well-known to be computationally challenging and numerically unstable when modeling wind at a large number of discrete points in space. Recently, stochastic waves have been used to model the field as a continuous process indexed both in time and space. We extend the classical Spectral Representation Method for simulation of Gaussian stochastic waves to a third-order representation modeling asymmetrically skewed non-Gaussian stochastic waves from a prescribed power spectrum and bispectrum. We present an efficient implementation using the fast Fourier transform, which reduces the computational time dramatically. We then apply the method for simulation of a non-Gaussian wind velocity field along a long-span bridge.
\end{abstract}


\end{frontmatter}

\section{Introduction}
\label{sec:introduction}


Dynamic wind loads can have unpredictable and devastating effects on the built environment. This is especially true for long-span structures where aerodynamic effects can cause instabilities, i.e.\ vortex-induced vibrations, galloping, and flutter.  Dynamic aeroelastic analysis is therefore critical to understanding the behavior of long-span structures (e.g.\ \citep{chen2000aerodynamic,Chen2010}) and these methods must account for the inherent randomness of the dynamic wind excitation. This is often done using Monte Carlo simulations, which require synthetic generation of the load-inducing random process \citep{Carassale2006}. Given this, simulation of stochastic wind fields is a topics of great interest in wind engineering. A wide range of approaches to simulate stochastic winds have been proposed (e.g.\ \cite{di1998digital,mann1998wind,cao2000simulation,li2004simplifying}) and many of them have been summarized by Kareem \cite{kareem2008numerical}.

Wind velocity fields are three-dimensional quantities with components in the $x, y$ and $z$ axes. That is, at any given point in space, $\mathbf{x}=\{x,y,z\}$, the wind velocity has three components denoted $\bm{u}_\mathbf{x}(t) = [u_\mathbf{x}(t), v_\mathbf{x}(t), w_\mathbf{x}(t)]^{T}$. However, prior investigations have shown that, although the three components are correlated, in practice the correlation may not be significant when compared to the overall uncertainty \cite{jones1992wind,minh1999numerical,oiseth2010simplified,Iseth2013}. This allows the three components to be modeled independently, which is common in engineering practice \cite{chen2000aerodynamic,strommen2010theory,tubino2007gust}. We therefore consider the simulation of a single scalar component, $u_\mathbf{x}(t)$, in this work.

Spatial correlation of a single velocity component, on the other hand, has been shown to be very important to the dynamic response of long-span structures.  The most common way to model spatially varying wind velocities is using a stochastic vector process \cite{deodatis1996simulation} $[u_{\mathbf{x}_1}(t), u_{\mathbf{x}_2}(t), \cdots, u_{\mathbf{x}_n}(t)]^{T}$ where each component of the vector process $u_{\mathbf{x}_i}(t)$ represents the scalar wind velocity at a single point ${\mathbf{x}_i}$ in space \cite{di2001digital}. While this approach works well for a small number of points ($n<  \sim10$), vector process simulation grows numerically unstable due to strong correlations between vector components causing singularities in the cross-spectral density matrix as $n$ grows larger. Although some approaches have been proposed to overcome these limitations \cite{ding2011efficient,gao2012improved,zhao2021simulation,li2022simulation}, a new way of looking at the problem was  proposed by Benowitz and Deodatis \citep{Benowitz2015}, who model spatially varying winds as stochastic waves instead of stochastic vector processes.  Methods for simulating stochastic waves were first proposed in the early 1990s by Shinozuka and Deodatis for seismic ground motions using the Spectral Representation Method \cite{shinozuka1991stochastic}, but have recently gained increased popularity -- with extensions e.g.\ to non-stationary \cite{peng2017simulation} and non-Gaussian \cite{Zhou2020} processes -- due to their ability to simulate across potentially very large spatio-temporal domains. The class of simulation methods proposed in \cite{shinozuka1991stochastic, Benowitz2015, peng2017simulation, Zhou2020} work by adapting the multi-dimensional version of the Spectral Representation Method to the time domain. This allows a random field, represented as $u(x)$, to be represented as a wave, $u(x, t)$. Apart from addressing the numerical instability discussed in \cite{ding2011efficient,gao2012improved,zhao2021simulation,li2022simulation}, the stochastic waves approach also eliminates the costly Cholesky decomposition step involved in simulating stochastic vector processes. This cost can become prohibitively large, particularly as the size of the spatial domain increases.


Turbulent wind velocity fields very often exhibit non-Gaussian properties \cite{Huang2016}. Methods for simulation of non-Gaussian stochastic waves are therefore essential for the accurate modeling of wind characteristics. While the literature for simulation of non-Gaussian stochastic processes is rich, there have been very few works that have extended these methods to stochastic waves. Methods for simulation of non-Gaussian processes include the direct transformation method \citep{Gurley1996}, the spectral correction method \citep{Masters2003}, translation process methods \cite{yamazaki1988digital,grigoriu1998simulation,gioffre2000simulation} including the Iterative Translation Approximation Method \citep{Shields2011,shields2013simple,Li2022} and moment/PDF-based Hermite polynomial translations \cite{yang2013probabilistic,yang2015efficient,puig2002non}, polynomial chaos expansions \citep{Sakamoto2002}, the Karhunen-Lo{\`{e}}ve expansion method \citep{Li2007,kim2015modeling}, the Johnson transformation \citep{Ma2016} and the tensor train Karhunen-Lo{\`{e}}ve \citep{Bu2020}. Among these many methods, only the ITAM has been extended for simulation of non-Gaussian stochastic waves \citep{Zhou2020}.


All the aforementioned methods are derived from second-order spectral properties, and are then transformed in some way to match higher-order moments or marginal distributions. In recent years, a new simulation framework has been developed that is derived from a higher-order expansion of the stochastic process \cite{Shields2017}. This method, which generalizes the widely-used Spectral Representation Method (SRM) for Gaussian processes \cite{Shields2017, vandanapu2022simulation, Vandanapu2021} is capable of matching the bispectrum (equivalently 3-point correlation) of the process in addition to the power spectrum (2-point correlation) that is standard. Very recently, Vandanapu et al.\ extended the $3^{rd}$-order SRM for simulation of non-stationary and non-Gaussian processes \cite{vandanapu2022simulation}, stochastic fields and vector processes \cite{Vandanapu2021}. Moreover, they introduced an efficient Fast Fourier Transform (FFT)-based implementation that drastically reduces computational cost. In this work, we extend the $3^{rd}$-order SRM to simulate non-Gaussian stochastic waves with a prescribed power spectrum and bispectrum. To improve the computational efficiency  of the simulation formula, we also introduce an FFT-based implementation. We then demonstrate this method for the simulation of a stochastic wind field approaching a long-span bridge structure.


\section{Properties of Stochastic Waves}

We begin by briefly reviewing the properties of stochastic waves. For the sake of simplicity, properties of 1-dimensional stochastic waves are presented but these properties can easily be extended to multi-dimensional waves. For a stationary, zero-mean, one-dimensional, homogeneous stochastic wave $u(x, t)$, the $2^{nd}$- and $3^{rd}$-order autocorrelation functions are defined as
\begin{equation}
\begin{aligned}
	& \mathbb{E}[u(x, t)] = 0\\
	& \mathbb{E}[u(x + \xi_{x}, t + \tau)u(x, t)] = R(\xi_{x}, \tau)\\
	& \mathbb{E}[u(x + \xi_{x1}, t + \tau_{1})u(x + \xi_{x2}, t + \tau_{2})u(x, t)] = R_{3}(\xi_{x1}, \xi_{x2}, \tau_{1}, \tau_{2})
\end{aligned}
\end{equation}
where $\xi_{x}$ is a spatial separation and $\tau$ is the time separation (or time lag). The power spectrum of the stochastic wave can be derived from its $2^{nd}$-order correlation function using the Weiner-Khintchine transform as
\begin{equation}
\begin{aligned}
	& S(\kappa_{x}, \omega) = \frac{1}{(2\pi)^{2}}\int_{-\infty}^{\infty}\int_{-\infty}^{\infty}R(\xi_{x}, \tau)e^{-\iota(\kappa_{x}\xi_{x} + \omega\tau)}d\xi_{x}d\tau
	\label{eqn:second_order_forward_weiner_khintchine}
\end{aligned}
\end{equation}
and the inverse of transform also exists as follows:
\begin{equation}
\begin{aligned}
	&R(\xi_{x}, \tau) = \int_{-\infty}^{\infty}\int_{-\infty}^{\infty} S(\kappa_{x}, \omega)e^{\iota(\kappa_{x}\xi_{x} + \omega\tau)}d\kappa_{x}d\omega
	\label{eqn:second_order_inverse_weiner_khintchine}
\end{aligned}
\end{equation}

Similarly, the bispectrum of a stochastic wave can be obtained from the $3^{rd}$-order correlation function as follows,
\begin{equation}
\begin{aligned}
	B(\kappa_{x1}, \kappa_{x2}, \omega_{1}, \omega_{2}) = & \frac{1}{(2\pi)^4}\int_{-\infty}^{\infty}\int_{-\infty}^{\infty}\int_{-\infty}^{\infty}\int_{-\infty}^{\infty}R_{3}(\xi_{x1}, \xi_{x2}, \tau_{1}, \tau_{2})\\
	& e^{-\iota(\kappa_{x1}\xi_{x1} + \kappa_{x2}\xi_{x2} + \omega_{1}\tau_{1} + \omega_{2}\tau_{2})} d\xi_{x1}d\xi_{x2}d\tau_{1}d\tau_{2}
	\label{eqn:third_order_forward_weiner_khintchine}
\end{aligned}
\end{equation}
where, again the inverse can be found as
\begin{equation}
\begin{aligned}
	R_{3}(\xi_{x1}, \xi_{x2}, \tau_{1}, \tau_{2}) = & \int_{-\infty}^{\infty}\int_{-\infty}^{\infty}\int_{-\infty}^{\infty}\int_{-\infty}^{\infty}B(\kappa_{x1}, \kappa_{x2}, \omega_{1}, \omega_{2})\\
	& e^{\iota(\kappa_{x1}\xi_{x1} + \kappa_{x2}\xi_{x2} + \omega_{1}\tau_{1} + \omega_{2}\tau_{2})} d\kappa_{x1}d\kappa_{x2}d\omega_{1}d\omega_{2}
	\label{eqn:third_order_inverse_weiner_khintchine}
\end{aligned}
\end{equation}

For wind fields, we are interested in the simulation of real-valued stochastic waves where the power spectrum is symmetric about the origin, i.e.
\begin{equation}
    S(\kappa_x, \omega) = S(-\kappa, \omega)
\end{equation}
and the bispectrum has the following of symmetries \cite{Shields2017}
\begin{equation}
\begin{aligned}
    & B(\kappa_{x_1}, \kappa_{x_2}, \omega_1, \omega_2) = B(\kappa_{x_2}, \kappa_{x_1}, \omega_2, \omega_1) \\
    & B(\kappa_{x_1}, \kappa_{x_2}, \omega_1, \omega_2) = B(-\kappa_{x_1}, -\kappa_{x_2}, -\omega_1, -\omega_2) \\
    & B(\kappa_{x_1}, \kappa_{x_2}, \omega_1, \omega_2) = B(-\kappa_{x_1} -\kappa_{x_2}, \kappa_{x_2}, -\omega_1 -\omega_2, \omega_2)
\end{aligned}
\end{equation}

Exploiting these symmetries, the power spectrum $S(\kappa_x, \omega)$ defined in the region $(-\infty < \kappa_{x} < \infty), (-\infty < \omega < \infty)$ can be sufficiently described as $2S(\kappa_x, \omega)$ in the region $(0 < \kappa_{x} < \infty), (-\infty < \omega < \infty)$. Similarly, the bispectrum $B(\kappa_{x_1}, \kappa_{x_2}, \omega_1, \omega_2)$ defined on the region $(-\infty < \kappa_{x1} < \infty, -\infty < \kappa_{x2} < \infty, -\infty < \omega_{1} < \infty, -\infty < \omega_{2} < \infty)$ can be replaced by $4B(\kappa_{x_1}, \kappa_{x_2}, \omega_1, \omega_2)$ on the region $(0 \leq \kappa_{x1} < \infty, 0 \leq \kappa_{x2} < \infty, -\infty < \omega_{1} < \infty, -\infty < \omega_{2} < \infty)$. These symmetries will be used in formulating fast and efficient simulation formulas.

\section{$2^{nd}$-Order Spectral Representation Method for Stochastic Waves}

Simulation of a stochastic wave $u(x, t)$ having frequency-wavenumber spectrum (FKS) $S(\kappa, \omega)$ according to the 2nd-order SRM is performed by evaluating the following summation of cosines \cite{Benowitz2015}
\begin{equation}
\begin{aligned}
	u(x, t) &= 2 \sum_{k_{\kappa}=0}^{N_{\kappa}-1}\sum_{k_{\omega}=0}^{N_{\omega}-1}\sum_{I_{\omega}=\pm} \sqrt{S(\kappa_{k_{\kappa}}, I_\omega \omega_{k_{\omega}})\Delta\kappa\Delta\omega}\cos(\kappa_{k_{\kappa}}x + I_\omega \omega_{k_{\omega}}t + \phi_{k_{\kappa}k_{\omega}}^{I_{\omega}}) 
\end{aligned}
\end{equation}
where
\begin{equation}
    \omega_{k_{\omega}} = k_{\omega}\Delta\omega; \Delta\omega = \frac{\omega_u}{N_{\omega}}; \kappa_{k_{\kappa}} = k_{\kappa}\Delta\kappa; \Delta\kappa = \frac{\kappa_{u}}{N_{\kappa}}
\end{equation}
and $\omega_u$, $\kappa_{u}$ are the frequency, and wavenumber cutoffs respectively, and $\phi_{k_{\kappa}k_{\omega}}^{I_{\omega}}$ are uniformly distributed random phase angles defined on $[0, 2\pi]$. The above formula can be rearranged to take advantage of the FFT technique as follows
\begin{equation}
    u(x, t) = \mathbb{Re} \Big[\sum_{k_{\kappa}=0}^{N_{\kappa} - 1} \sum_{k_{\omega}=0}^{N_\omega - 1} [B^{(+)}_{k_{\kappa}k_{\omega}}\exp(\iota \omega_{k_{\omega}} t + \iota \kappa_{k_{\kappa}} x) + B^{(-)}_{k_{\kappa}k_{\omega}}\exp(-\iota \omega_{k_{\omega}} t + \iota \kappa_{k_{\kappa}} x)]\Big]
\label{eqn:2_order_fft}
\end{equation}
where $\iota$ is the imaginary unit, $\mathbb{Re}[\cdot]$ denotes the real part and
\begin{equation}
    B^{(\pm)}_{k_{\kappa}k_{\omega}} = 2 \sqrt{S(\kappa_{k_\kappa}, \pm \omega_{k_\omega})\Delta \omega \Delta \kappa} \cdot \exp(\iota \phi_{\kappa_{k_\kappa}  \omega_{k_\omega}}^{(\pm)})
\end{equation}
where $\phi^{(n)}$ are sets of independent uniform random variables on $[0, 2\pi]$ of the shape $[N_\kappa, N_\omega]$. Eq[\ref{eqn:2_order_fft}] can be represented in a compact form as
\begin{equation}
    u(x, t) = \mathbb{Re} \Big[ \texttt{FFT}_{}[\texttt{FFT}_{}(B_{k_\kappa k_\omega }^{(+)})] + \texttt{FFT}_{}[\texttt{IFFT}_{}(B_{k_\kappa k_\omega }^{(-)})] \Big]
\end{equation}
where $\texttt{FFT}$ and $\texttt{IFFT}$ denote Fast Fourier Transform and Inverse Fast Fourier Transform respectively.

\section{$3^{rd}$-Order Spectral Representation Method for Stochastic Waves}

Shields and Kim \cite{Shields2017} first proposed the simulation of non-Gaussian stochastic processes using the $3^{rd}$-order Spectral Representation of the following form
\begin{equation}
\begin{aligned}
    u(t) = & 2 \Bigg[ \sum_{k=0}^{\infty} \sqrt{S_p(\omega_k)\Delta\omega_k} \cos(\omega_k t + \phi_k) + \\
    & \sum_{i + j = k}^{i \leq j \leq 0} \sqrt{S(\omega_k)\Delta\omega_k} |b_{p}(\omega_i, \omega_j)| \cos(\omega_k t + \phi_i + \phi_j + \beta(\omega_i, \omega_j)) \Bigg]
\end{aligned}
\end{equation}
where $S(\omega)$ is the power spectrum, $S_p(\omega)$ is the so-called pure power spectrum defined as
\begin{equation}
    S_p(\omega_k) = S(\omega_k)(1 - \sum_{i + j = k}^{i \leq j \leq 0} b_p^2(\omega_i, \omega_j)),
    \end{equation}
$b_p(\omega_i, \omega_j)$ is the partial bicoherence defined as
\begin{equation}
    b_p^2(\omega_i, \omega_j) = \frac{|B(\omega_i, \omega_j)|^2 \Delta \omega_i^2 \Delta \omega_j^2}{S_p(\omega_i)\Delta \omega_i S_p(\omega_j) \Delta \omega_j S(\omega_i + \omega_j) \Delta (\omega_i + \omega_j)},
\end{equation}
$\beta(\omega_i, \omega_j)$ is the biphase defined as
\begin{equation}
    \beta(\omega_i, \omega_j) = \arctan \left\{ \frac{\Im[B(\omega_i, \omega_j)]}{\Re[B(\omega_i, \omega_j)]} \right\},
\end{equation}
and $\Delta \omega$ is the frequency interval
\begin{equation}
    \Delta \omega = \frac{\omega_u}{N},
\end{equation}
$\omega_u$ is the cutoff frequency, $\omega_k = k \Delta \omega$ and $\phi_k$ are uniformly distributed random phases angle on $[0, 2\pi]$. An FFT based implementation, not shown here, was derived in \cite{Vandanapu2021}. 

\subsection{Simulation of General 1-dimensional Stochastic Waves}

The original $3^{rd}$-order simulation formula was derived by defining a set of orthogonal increments in the original spectral representation that satisfy requisite higher-order orthogonality conditions \cite{Shields2017,Vandanapu2021}. Likewise, it's straightforward to derive an expansion for the simulation of stochastic waves in space-time by specifying such orthogonal increments. The details are not shown here due to their length and similarity to those derived in \cite{Shields2017,Vandanapu2021}. Instead, focus simply on the resulting simulation formulas that enable stochastic wave generation for wind field simulations.

Let $u(x, t)$ be a one-dimensional stochastic wave with zero mean, FKS $S(\kappa, \omega)$ and bispectrum $B(\kappa_1, \kappa_2, \omega_1, \omega_2)$. Simulation of the stochastic wave $u(x, t)$ can be performed through the following summation
\begin{equation}
\begin{aligned}
	u(x, t) = & 2 \sum_{k_\kappa=0}^{N_\kappa} \sum_{k_\omega=-N\omega}^{N_\omega} \Bigg[ \sqrt{S_{p}(\kappa_{k_\kappa}, \omega_{k_\omega}) \Delta\kappa \Delta \omega} \cos(\kappa_{k_\kappa}x + \omega_{k_\omega}t + \phi_{k_\kappa k_\omega}) +\\
& \sum_{i_\kappa + j_\kappa = k_\kappa }^{i_\kappa \geq j_\kappa \geq 0} \sum_{i_\omega + j_\omega = k_\omega}^{N_{\omega} \geq |i_\omega| \geq |j_\omega| \geq 0} \sqrt{S(\kappa_{k_{\kappa}}, \omega_{k_\omega})\Delta\kappa\Delta\omega}  b(\kappa_{i_\kappa}, \kappa_{j_\kappa}, \omega_{i_\omega}, \omega_{j_\omega})\\
& \cos(\kappa_{k_\kappa}x + \omega_{k_\omega}t + \phi_{i_\kappa i_\omega} + \phi_{j_\kappa j_\omega}  + \beta(\kappa_{i_\kappa}, \kappa_{j_\kappa}, \omega_{i_\omega}, \omega_{j_\omega}) ) \Bigg]
\label{eqn:truncated_cosines_1d_wave}
\end{aligned}
\end{equation}
where $\Delta \kappa$ and $\Delta \omega$ are the wave-number interval and frequency intervals defined as
\begin{equation}
\begin{aligned}
    &\Delta \kappa = \frac{\kappa_{u}}{N_\kappa}\\
	&\Delta \omega = \frac{\omega_{u}}{N_{\omega}}\\
\end{aligned}
\end{equation}
where $\kappa_{u}$ is the cutoff wave-number and $\omega_u$ is the cutoff frequency. The indexed wave-number and frequency are given as
\begin{equation}
\begin{aligned}
    & \kappa_{k_\kappa} = k_\kappa * \Delta \kappa \\
    & \omega_{k_\omega} = k_\omega * \Delta \omega \\
\end{aligned}
\end{equation}
The pure component of the FKS $S_p(\kappa, \omega)$ is defined as
\begin{equation}
    S_p(\kappa_{k_\kappa}, \omega_{k_\omega}) = S(\kappa_{k_\kappa}, \omega_{k_\omega})\left(1 - \sum_{i_\kappa + j_\kappa = k_\kappa }^{i_\kappa \geq j_\kappa \geq 0} \sum_{i_\omega + j_\omega = k_\omega}^{N_{\omega} \geq |i_\omega| \geq |j_\omega| \geq 0} b_{p}(\kappa_{i_\kappa}, \kappa_{j_\kappa}, \omega_{i_\omega}, \omega_{j_\omega}) \right),
    \label{eqn:pure}
\end{equation}
the partial bicoherence is defined as
\begin{equation}
    b_{p}(\kappa_{i_\kappa}, \kappa_{j_\kappa}, \omega_{i_\omega}, \omega_{j_\omega}) = \frac{|B(\kappa_{i_\kappa}, \kappa_{j_\kappa}, \omega_{i_\omega}, \omega_{j_\omega})|\sqrt{\Delta \kappa \Delta \omega}}{\sqrt{S_{p}(\kappa_{i_\kappa}, \omega_{i_\omega})S_{p}(\kappa_{j_\kappa}, \omega_{j\omega})S(\kappa_{k_\kappa}, \omega_{k_\omega})}}
\end{equation}
and the biphase is defined as
\begin{equation}
    \beta(\kappa_{i_\kappa}, \kappa_{j_\kappa}, \omega_{i_\omega}, \omega_{j_\omega}) = \arctan \left\{ \frac{\Im[B(\kappa_{i_\kappa}, \kappa_{j_\kappa}, \omega_{i_\omega}, \omega_{j_\omega})]}{\Re[B(\kappa_{i_\kappa}, \kappa_{j_\kappa}, \omega_{i_\omega}, \omega_{j_\omega})]} \right\}.
\end{equation}
Finally, $\phi_{k_\kappa k_\omega}$ are uniformly distributed phase angles distributed in $[0, 2\pi]$.

\subsection{Simulation of Quadrant 1-dimensional Stochastic Waves}

Qaudrant stochastic waves have symmetries along all axes. The symmetry equations for a 1-dimensional stochastic wave are given by
\begin{equation}
\begin{aligned}
	&S(\kappa, \omega) = S(I_{\kappa}\kappa, I_{\omega}\omega) \ \text{for} \ I_{\kappa},  I_{\omega} = \pm 1\\
	&B(\kappa_{1}, \kappa_{2}, \omega_{1}, \omega_{2}) = B(I_{\kappa_1}\kappa_{1}, I_{\kappa_2}\kappa_{2}, I_{\omega_1}\omega_{1}, I_{\omega_2}\omega_{2}) \ \text{for} \ I_{\kappa_1},  I_{\kappa_2}, I_{\omega_1}, I_{\omega_2} = \pm 1
\end{aligned}
\end{equation}

Leveraging these additional symmetries, the simulation formula for 1-dimensional stochastic wave can be rewritten as
\begin{equation}
\begin{aligned}
    u(x, t) = & 2 \sum_{k_\kappa=0}^{N_\kappa}\sum_{k_\omega=0}^{N_\omega}\sqrt{S_{p}(\kappa_{k_\kappa}, \omega_{k_\omega}) \Delta\kappa\Delta\omega} \Bigg[\cos(\kappa_{k_\kappa}x + \omega_{k_\omega} t + \phi_{k_\kappa k_\omega}^{(+)})\\
    & + \cos(\kappa_{k_\kappa}x - \omega_{k_\omega} t + \phi_{k_\kappa k_\omega}^{(-)})\\
    & + \sum_{i_\kappa + j_\kappa = k_\kappa}^{i_\kappa \geq j_\kappa \geq 0}\sum_{_\omega + j_\omega = k_\omega}^{i_\omega \geq j_\omega \geq 0}\sqrt{S(\kappa_{k_\kappa}, \omega_{k_\omega}) \Delta\kappa \Delta\omega } b_{p}(\kappa_{i_\kappa}, \kappa_{j_\kappa}, \omega_{i_\omega}, \omega_{j_\omega})\\
    &\Big[ \cos(\kappa_{k_\kappa}x + \omega_{k_\omega}t + \phi_{i_\kappa i_\omega}^{(+)} + \phi_{j_{x}j}^{(+)} + \beta(\kappa_{i_{x}}, \kappa_{j_{x}}, \omega_{i}, \omega_{j}))\\
    & + \cos(\kappa_{k_\kappa}x - \omega_{k_\omega}t + \phi_{i_\kappa i_\omega}^{(-)} + \phi_{j_\kappa j_\omega}^{(-)} + \beta(\kappa_{i_\kappa}, \kappa_{j_\kappa}, \omega_{i_\omega}, \omega_{j_\omega})) \\
   & + \cos(\kappa_{k_\kappa}x - \omega_{i_\omega}t + \omega_{j_\omega}t + \phi_{i_\kappa i_\omega}^{(-)} + \phi_{j_\kappa j_\omega}^{(+)} + \beta(\kappa_{i_\kappa}, \kappa_{j_\kappa}, \omega_{i_\omega}, \omega_{j_\omega}))\\
   & + \cos(\kappa_{k_\kappa}x + \omega_{i_\omega}t - \omega_{j_\omega}t + \phi_{i_\kappa i_\omega}^{(+)} + \phi_{j_\kappa j_\omega}^{(-)} + \beta(\kappa_{i_\kappa}, \kappa_{j_\kappa}, \omega_{i_\omega}, \omega_{j_\omega})) \Big] \Bigg]
\label{eqn:quadrant_1d_wave}
\end{aligned}
\end{equation}
where the terms of the equation are the same as in Eq.\ \eqref{eqn:truncated_cosines_1d_wave}. The phase angles $\phi^{(+)}$ correspond to uniformly distributed phase angles for $\omega > 0$ and $\phi^{(-)}$ corresponds to $\omega < 0$. Although the form of the equation looks more complicated, this form of the equation is easier to implement because of the symmetries.

\subsection{FFT implementation}

The simulation formula as shown above is computationally very expensive. Implementing the formula using FFT can dramatically improve the computational efficiency. The simulation formula defined in Eq.\ \eqref{eqn:quadrant_1d_wave} can be re-written as

\begin{equation}
\begin{aligned}
    u(m_{x} \Delta x, m_{t} \Delta t) = & 2 \sum_{k_{\kappa}=0}^{N_{\kappa}}\sum_{k_\omega=0}^{N_{\omega}}\sqrt{S(k_{\kappa} \Delta \kappa, k_\omega \Delta \omega) \Delta \kappa \Delta \omega} \\
    & \Big[ \sqrt{(1 - \sum_{i_\kappa + j_\kappa = k_\kappa}^{i_\kappa \geq j_\kappa \geq 0}\sum_{i_\omega + j_\omega = k_\omega}^{N_\omega \geq |i_\omega| \geq |j_\omega| \geq 0} b_{p}^{2}(i_\kappa \Delta \kappa,  j_\kappa \Delta \kappa, i_\omega \Delta \omega, j_\omega \Delta \omega)} e^{\iota\phi_{k_\kappa k_\omega}^{(+)}}\\
    & + [\sum_{i_\kappa + j_\kappa = k_\kappa}^{i_\kappa \geq j_\kappa \geq 0}\sum_{i_\omega + j_\omega = k_\omega}^{N_\omega \geq |i_\omega| \geq |j_\omega| \geq 0} b_{p}(i_\kappa \Delta \kappa, j_\kappa \Delta \kappa, i_\omega \Delta \omega, j_\omega \Delta \omega)\\
    & e^{\iota(\Phi_{i_\kappa i_\omega}^{(+|-)} + \Phi_{j_\kappa j_\omega}^{(+|-)} +\beta(i_\kappa \Delta \kappa, j_\kappa \Delta \kappa, i_\omega \Delta \omega, j_\omega \Delta \omega))}] \Big] e^{\iota(k_\kappa \Delta \kappa m_{x} \Delta x + k_\omega \Delta \omega m_{t} \Delta t)}\\
    & + 2 \sum_{k_{\kappa}=0}^{N_{\kappa}}\sum_{k_\omega=0}^{N_{\omega}}\sqrt{S(k_{\kappa} \Delta \kappa, k_\omega \Delta \omega) \Delta \kappa \Delta \omega} \\
    & \Big[ \sqrt{(1 - \sum_{i_\kappa + j_\kappa = k_\kappa}^{i_\kappa \geq j_\kappa \geq 0}\sum_{i_\omega + j_\omega = -k_\omega}^{N_\omega \geq |i_\omega| \geq |j_\omega| \geq 0} b_{p}^{2}(i_\kappa \Delta \kappa,  j_\kappa \Delta \kappa, i_\omega \Delta \omega, j_\omega \Delta \omega)} e^{\iota\phi_{k_\kappa k_\omega}^{(-)}}\\
    & + [\sum_{i_\kappa + j_\kappa = k_\kappa}^{i_\kappa \geq j_\kappa \geq 0}\sum_{i_\omega + j_\omega = -k_\omega}^{N_\omega \geq |i_\omega| \geq |j_\omega| \geq 0} b_{p}(i_\kappa \Delta \kappa, j_\kappa \Delta \kappa, i_\omega \Delta \omega, j_\omega \Delta \omega)\\
    & e^{\iota(\Phi_{i_\kappa i_\omega}^{(+|-)} + \Phi_{j_\kappa j_\omega}^{(+|-)} +\beta(i_\kappa \Delta \kappa, j_\kappa \Delta \kappa, i_\omega \Delta \omega, j_\omega \Delta \omega))}] \Big] e^{\iota(k_\kappa \Delta \kappa m_{x} \Delta x - k_\omega \Delta \omega m_{t} \Delta t)}
\label{eqn:fft_2d_2_parts}
\end{aligned}
\end{equation}
where $\Phi^{(+|-)}$ represents the appropriate phase angles, $\Phi^{(+)}$ or $\Phi^{(-)}$, selected as: $\Phi^{(+)}$ for $i_{\omega}$ or $j_{\omega} > 0$, and $\Phi^{(-)}$ for $i_{\omega}$ or $j_{\omega} < 0$. The use of FFT requires that the interval products follow the following relations
\begin{equation}
\begin{aligned}
	& \Delta \omega \Delta t = \frac{2 \pi}{N_t} \\
	& \Delta \kappa \Delta x = \frac{2 \pi}{N_x}.
\end{aligned}
\end{equation}
Eq.\ \eqref{eqn:fft_2d_2_parts} can then be expressed in the following form, on which Fast Fourier transform can be applied easily
\begin{equation}
\begin{aligned}
	&u(x, t) = \Re \Bigg[ \sum_{k_\omega=0}^{N_\omega} \sum_{k_\kappa=0}^{N_\kappa} B_{k_\kappa k_\omega}^{(+)} \exp(\iota \omega_{k_\omega}t + \iota \kappa_{k_\kappa} x) + B_{k_\kappa k_\omega}^{(-)} \exp(-\iota \omega_{k_\omega}t + \iota \kappa_{k_\kappa} x) \Bigg]
\end{aligned}
\end{equation}
where
\begin{equation}
\begin{aligned}
	B^{(+)}_{k_\kappa k_\omega } &= \sum_{k_\kappa=0}^{N_\kappa}\sum_{k_\omega=0}^{N_\omega} \sqrt{S(k_\kappa \Delta \kappa, k_\omega \Delta \omega) \Delta\kappa \Delta\omega}\\
    & \Big[ \sqrt{(1 - \sum_{i_\kappa + j_\kappa = k_\kappa }^{i_\kappa \geq j_\kappa \geq 0}\sum_{i_\omega + j_\omega = k_\omega }^{N_\omega \geq |i_\omega| \geq |j_\omega| \geq 0}b_{p}^{2}(i_\kappa \Delta \kappa, j_\kappa \Delta \kappa, i_\omega \Delta \omega, j_\omega \Delta \omega)} e^{\iota\Phi_{k_\kappa k_\omega}^{(+)}}\\
    & + [\sum_{i_\kappa + j_\kappa = k_\kappa}^{i_\kappa \geq j_\kappa \geq 0}\sum_{i_\omega + j_\omega = k_\omega}^{k_{\omega} \geq |i_{\omega}| \geq |j_{\omega}| \geq 0} b_{p}(i_\kappa \Delta \kappa, j_\kappa \Delta \kappa, i_{\omega} \Delta \omega, j_{\omega} \Delta \omega) e^{\iota(\Phi_{i_\kappa i_{\omega}}^{(+|-)} + \Phi_{j_\kappa j_{\omega}}^{(+|-)} +\beta(i_\kappa \Delta \kappa, j_\kappa \Delta \kappa, i_{\omega} \Delta \omega, j_{\omega} \Delta \omega))}] \Big]\\
\end{aligned}
\end{equation}
and
\begin{equation}
\begin{aligned}
	B^{(-)}_{k_\kappa k_\omega } &= \sum_{k_\kappa=0}^{N_\kappa}\sum_{k_\omega=0}^{N_\omega} \sqrt{S(k_\kappa \Delta \kappa, k_\omega \Delta \omega) \Delta\kappa \Delta\omega}\\
    & \Big[ \sqrt{(1 - \sum_{i_\kappa + j_\kappa = k_\kappa }^{i_\kappa \geq j_\kappa \geq 0}\sum_{i_\omega + j_\omega = -k_\omega }^{N_\omega \geq |i_\omega| \geq |j_\omega| \geq 0}b_{p}^{2}(i_\kappa \Delta \kappa, j_\kappa \Delta \kappa, i_\omega \Delta \omega, j_\omega \Delta \omega)} e^{\iota\Phi_{k_\kappa k_\omega}^{(-)}}\\
    & + [\sum_{i_\kappa + j_\kappa = k_\kappa}^{i_\kappa \geq j_\kappa \geq 0}\sum_{i_\omega + j_\omega = -k_\omega}^{k_{\omega} \geq |i_{\omega}| \geq |j_{\omega}| \geq 0} b_{p}(i_\kappa \Delta \kappa, j_\kappa \Delta \kappa, i_{\omega} \Delta \omega, j_{\omega} \Delta \omega) e^{\iota(\Phi_{i_\kappa i_{\omega}}^{(+|-)} + \Phi_{j_\kappa j_{\omega}}^{(+|-)} +\beta(i_\kappa \Delta \kappa, j_\kappa \Delta \kappa, i_{\omega} \Delta \omega, j_{\omega} \Delta \omega))}] \Big]\\
\end{aligned}
\end{equation}

Use of FFT reduces the time complexity  of the simulation formula from $O(N_\kappa N_\omega N_t N_x)$ to $O(N_\kappa N_\omega \log (N_t N_x))$. While the use of FFT reduces the computational complexity of the simulation, the memory requirement remains unchanged. The memory required for storing the bispectrum is of the order of $O(N_{\kappa}^{2}N_{\omega}^{2})$. Assuming the use of 64 bit floating point variable type, the memory required to store a bispectrum with 512 $\kappa$-intervals and 256 $\omega$-intervals is 125 GB. Meanwhile, the memory required with 1024 $\kappa$-intervals and 512 $\omega$-intervals (doubling the size of the problem) is 2 TB. Therefore, the scale of the memory requirement prohibits naively scaling the simulation procedure. This is the primary drawback of the proposed approach and practically limits the size of the spatio-temporal domain that can be realistically simulated. For large simulation domains, distributed storage and retrieval or just-in-time computation of required bispectrum values may help to scale the simulation formula. However, these methods have not been explored in this work and require further research.

Although the high-memory requirement $O(N_{\kappa}^{2}N_{\omega}^{2})$ (in the stationary case) might appear restrictive, it is still an improvement over the $O(N_{x}^{3}N_{\omega}^{2})$ requirement of stochastic vector process. This is particularly crucial when the spatial domain is large. As discussed Sec[\ref{sec:introduction}], the stochastic waves approach also eliminates the expensive Cholesky decomposition procedure at the simulation stage. Given these benefits, the stochastic waves approach is superior to the stochastic vector approach, especially when the spatial domain is large.

\section{Simulation of a $3^{rd}$-order Wind Field}

\begin{figure}[!ht]
    \centering
    \includegraphics{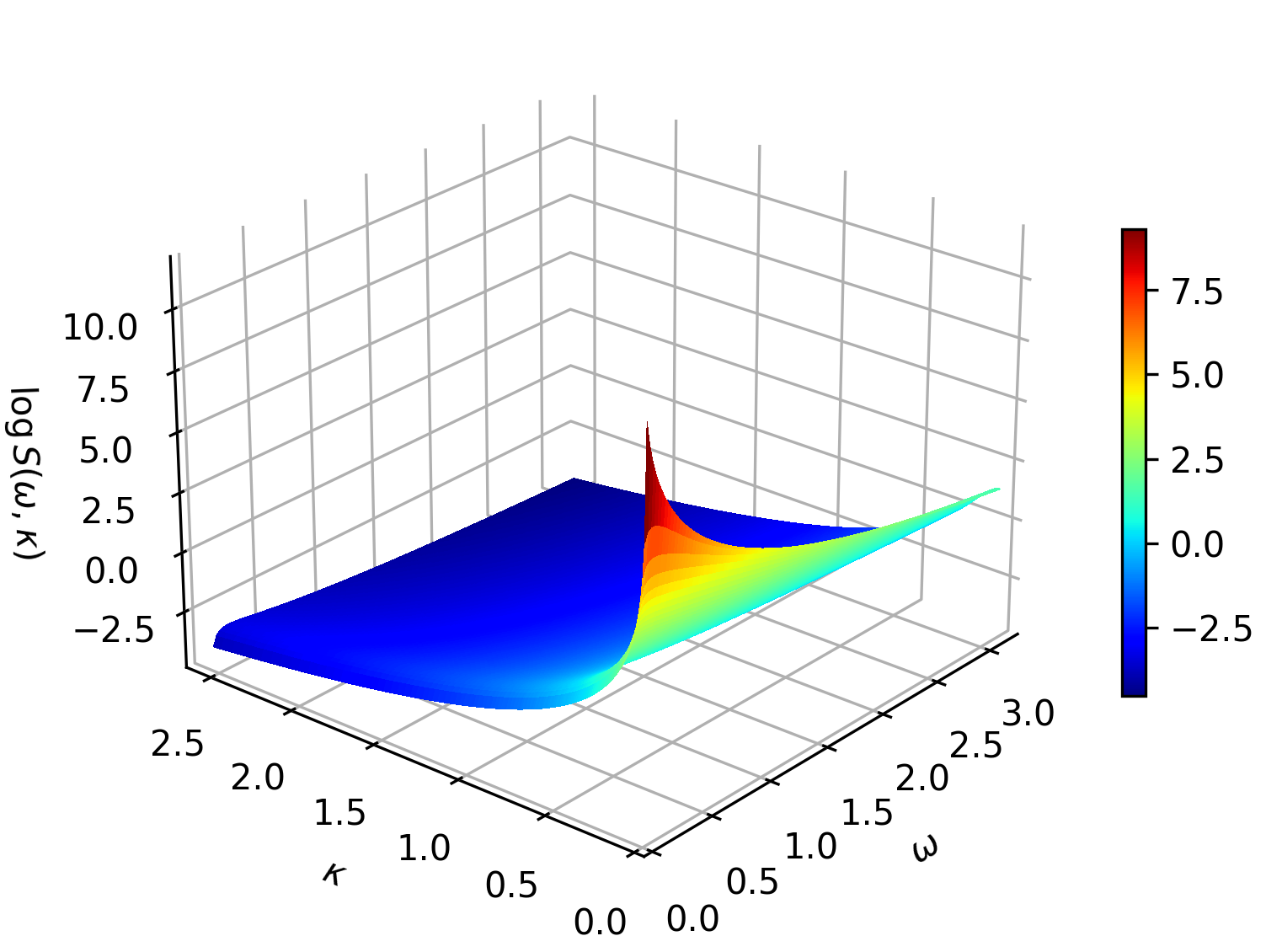}
    \caption{Frequency-Wavenumber Spectrum (FKS)}
    \label{fig:FKS_spectrum}
\end{figure}

In this section, we use the proposed approach to simulate stochastic wave wind velocities on a long-span bridge structure. The wind field is assumed to follow a Kaimal power spectrum \cite{Kaimal1972} given by
\begin{equation}
	S(\omega) = \frac{1}{2} \frac{200}{2\pi} u_{*}^{2} \frac{z}{U(z)} \frac{1}{\Big[1+50\frac{\omega z}{2\pi U(z)}\Big]^{5/3}}
	\label{eqn:kaimal_spectrum}
\end{equation}
where $z$ is the height above the ground, $U(z)$ is the average wind velocity at height $z$, and $u_{*}$ is the shear velocity defined as 
\begin{equation}
    u_{*} = \frac{kU(z)}{ln( z / z_{0})}
\end{equation}
where $k$ is the von Karman constant and $z_0$ describes the ground roughness. The parameter values are assigned as
\begin{equation}
\begin{aligned}
    & z = 50m \\
    & z_0 = 0.03 \\
    & U(z) = 40 m/s\\
    & k = 0.4 \\
\end{aligned}
\end{equation}
The target coherence is given by the Davenport \cite{Davenport1967} coherence function
\begin{equation}
    \gamma(\xi, \omega) = \exp \Big[ - \frac{\lambda \omega \xi}{2 \pi U(z)} \Big]
    \label{eqn:davenport_coherence}
\end{equation}
with coherence parameter $\lambda = 10$. The target FKS can then be defined as
\begin{equation}
    S(\omega, \kappa) = \frac{1}{2\pi} \int_{-\infty}^{\infty} S(\omega)\gamma(\xi, \omega) e^{\iota \xi \kappa} d\xi
    \label{eqn:fks_integral}
\end{equation}
Substituting Eqs. (\ref{eqn:kaimal_spectrum}, \ref{eqn:davenport_coherence}) into Eq. \eqref{eqn:fks_integral} and solving the integral yields \cite{Zhou2020}
\begin{equation}
    S(\omega, \kappa) = \frac{1}{2} \frac{200}{2\pi} u_{*}^{2} \frac{z}{U(z)} \frac{1}{\Big[1+50\frac{\omega z}{2\pi U(z)}\Big]^{5/3}} \Bigg[\frac{\Big(\frac{\lambda \omega}{\pi U(z)}\Big)}{\kappa^{2} + \Big(\frac{\lambda \omega}{\pi U(z)}\Big)^{2}}\Bigg]
    \label{eqn:kaimal_davenport_spectrum}
\end{equation}
The length of the bridge is $L=628m$. The cutoff frequency for simulation is chosen to be $\omega_u = \pi$ rad/sec and the frequency, time, space, and wave number intervals are assigned as follows
\begin{equation}
\begin{aligned}
    & N_\omega = 128; \quad \Delta \omega = \frac{\omega_u}{N_\omega - 1} = 0.0247 \text{ rad/sec} \\
    & N_t = 2048; \quad T = \frac{2\pi}{\Delta \omega} = 254 \text{ sec}; \quad \Delta t = \frac{T}{N_t} = 0.124 \text{ sec} \\
    & N_x = 1024; \quad \Delta x  = \frac{L}{N_x} = 0.6135 \text{ m} \\
    & N_\kappa = 256; \quad \Delta \kappa = \frac{2\pi}{L} = 0.001 \text{ m}^{-1}; \quad \kappa_u = \frac{2(N_\kappa - 1)\pi}{L} = 2.55 \text{ m}^{-1}
\end{aligned}
\end{equation}
The FKS having the above parameters is plotted in Fig.\ \ref{fig:FKS_spectrum} and results in a process having standard deviation 18.45 m/sec (variance 340.41, see Table \ref{tab:statistics_table}).

The bispectrum for the wind field is chosen as
\begin{equation}
    B(\omega_1, \omega_2, \kappa_1, \kappa_2) = \frac{2\sqrt{S(\omega_1, \kappa_1) S(\omega_2, \kappa_2)S(\omega_1 + \omega_2, \kappa_1 + \kappa_2)}}{3\sqrt{3(\omega_1 + \omega_2)(\kappa_1 + \kappa_2)}}
    \label{eqn:bispectrum}
\end{equation}
where $S$ is the Kaimal power spectrum defined in \ref{eqn:kaimal_davenport_spectrum}. Due to the four-dimensional nature of this function, it is impossible to plot this function for visualization. However, it can be shown that the process has skewness 0.497 (see Table \ref{tab:statistics_table}). The selection of bispectrum here is arbitrary but has been defined to ensure it possesses certain characteristics. In particular, it is defined as a well-behaved function (having strictly positive pure power spectrum, Eq.\ \eqref{eqn:pure}) of the power spectrum having positive skewness.  



\begin{figure}[!ht]
\centering
\begin{subfigure}{0.49\textwidth}
  \includegraphics[width=0.8\linewidth]{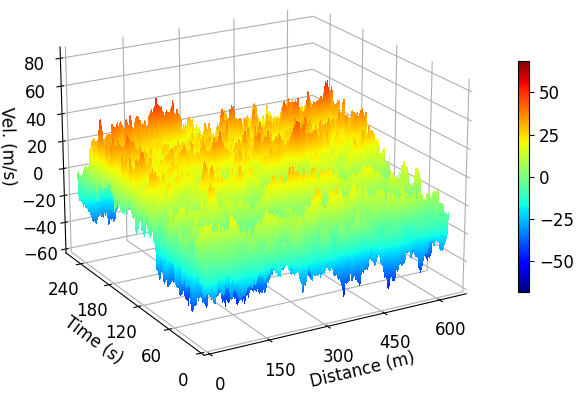}
  \label{fig:sample_srm}
  \caption{}
 \end{subfigure}
 \begin{subfigure}{0.49\textwidth}
  \includegraphics[width=0.8\linewidth]{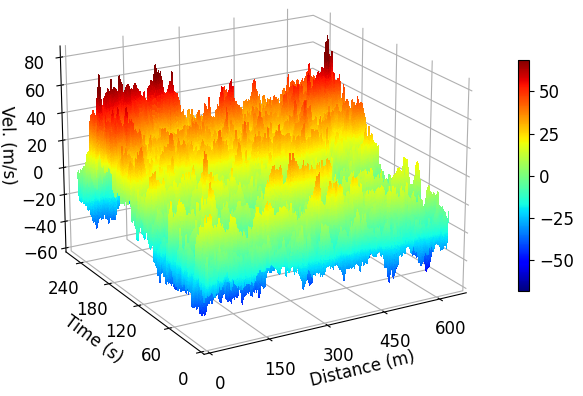}
  \label{fig:sample_bsrm}
  \caption{}
 \end{subfigure}
  \begin{subfigure}{0.50\textwidth}
  \includegraphics[width=0.8\linewidth]{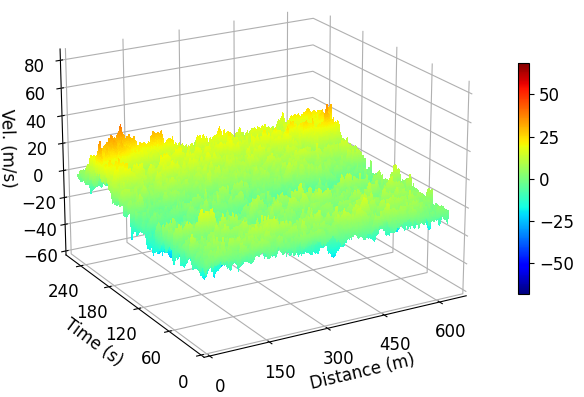}
  \label{fig:sample_diff}
  \caption{}
 \end{subfigure}
\caption{Sample realization of a wind field generated as a stochastic wave using (a) 2nd-order Spectral Representation Method and (b) 3rd-order Spectral Representation Method. The difference between these realizations is shown in (c).}
\label{fig:sample_plots}
\end{figure}

\begin{figure}[!ht]
    \centering
    \begin{subfigure}{0.49\textwidth}
        \begin{subfigure}{\textwidth}
        \includegraphics[width=1.0\linewidth]{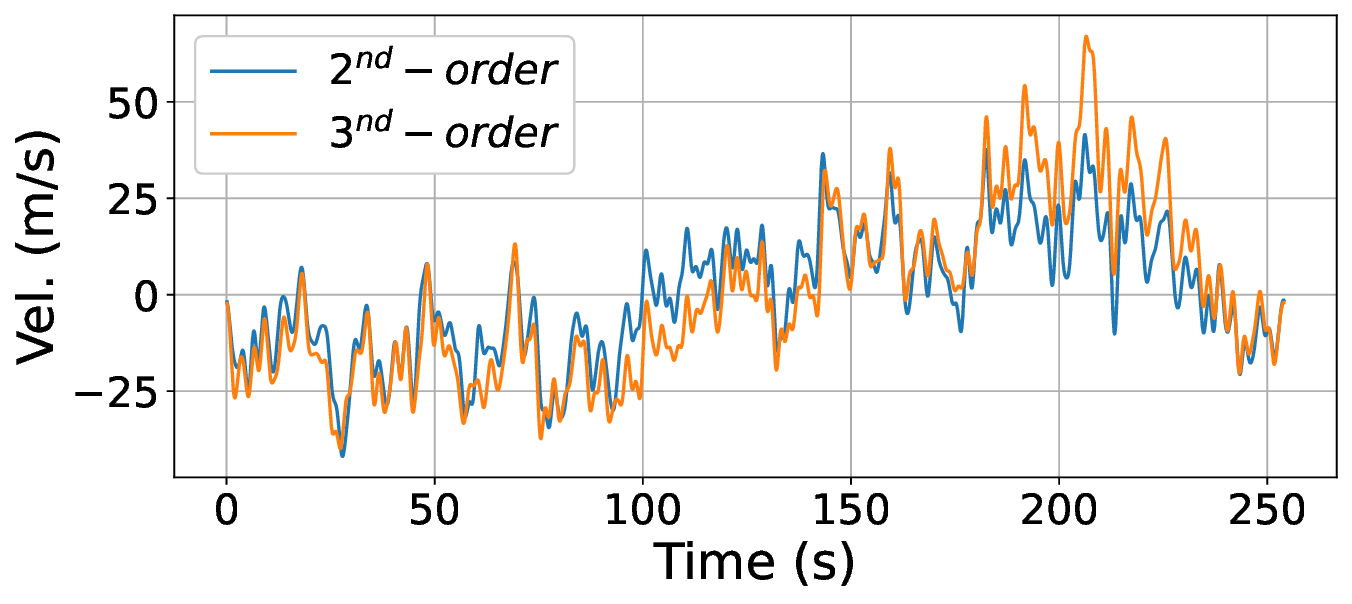}
        \caption{$x=0$ m.}
        \end{subfigure}
        \begin{subfigure}{\textwidth}
        \includegraphics[width=1.0\linewidth]{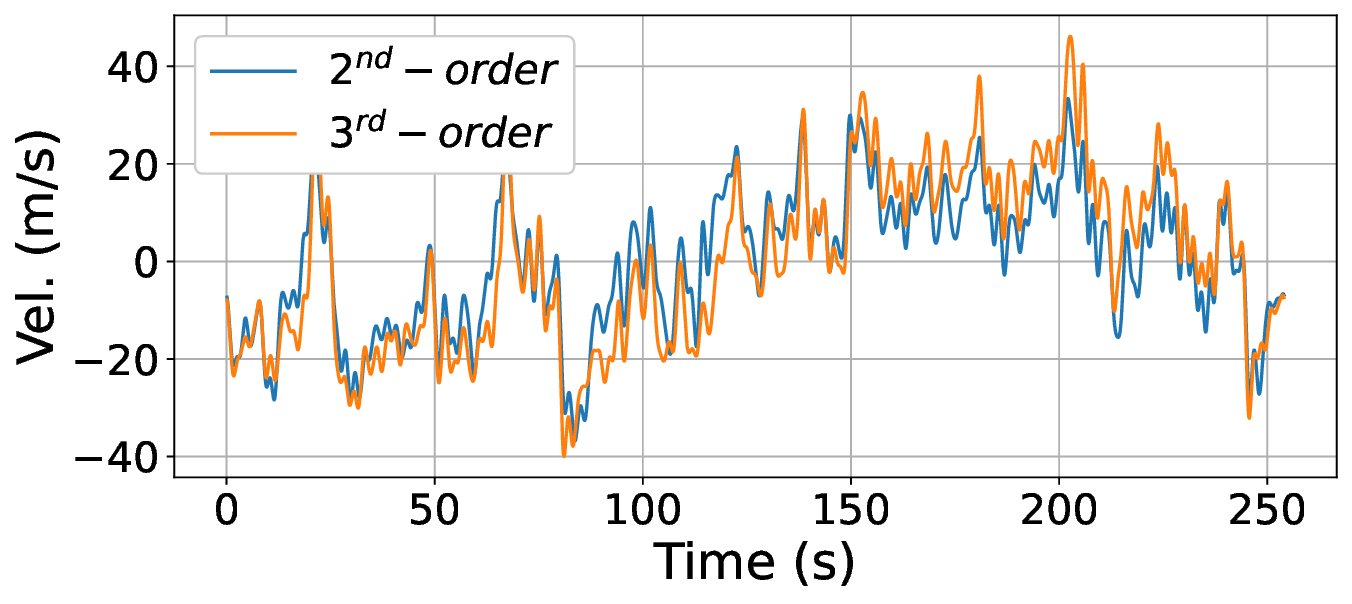}
        \caption{$x=314$ m.}
        \end{subfigure}
        \begin{subfigure}{\textwidth}
        \includegraphics[width=1.0\linewidth]{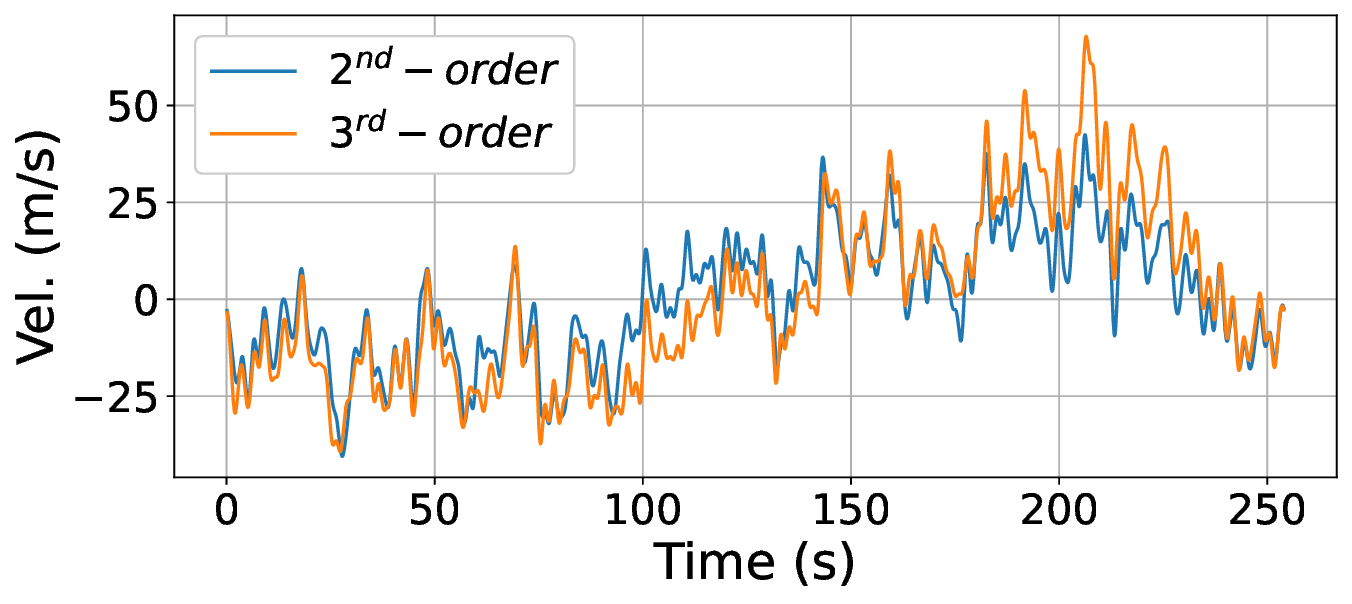}
        \caption{$x=628$ m.}
        \end{subfigure}
    \end{subfigure}
    \begin{subfigure}{0.49\textwidth}
        \begin{subfigure}{\textwidth}
        \includegraphics[width=1.0\linewidth]{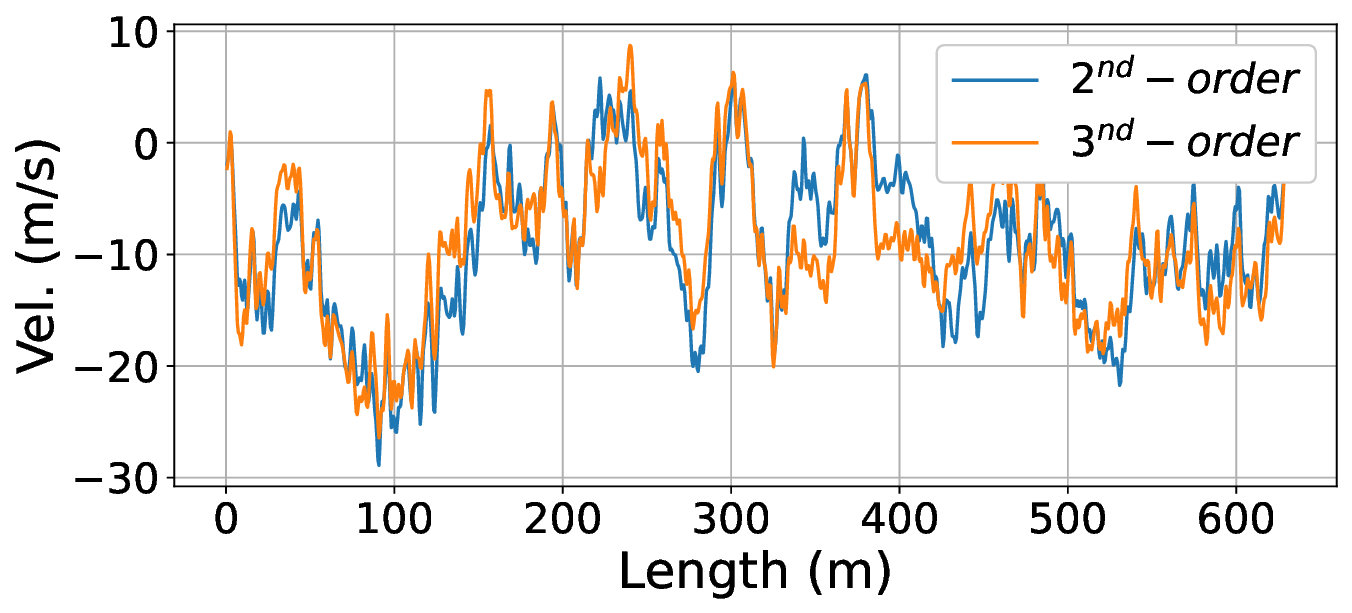}
        \caption{$t=0$ sec.}
        \end{subfigure}
        \begin{subfigure}{\textwidth}
        \includegraphics[width=1.0\linewidth]{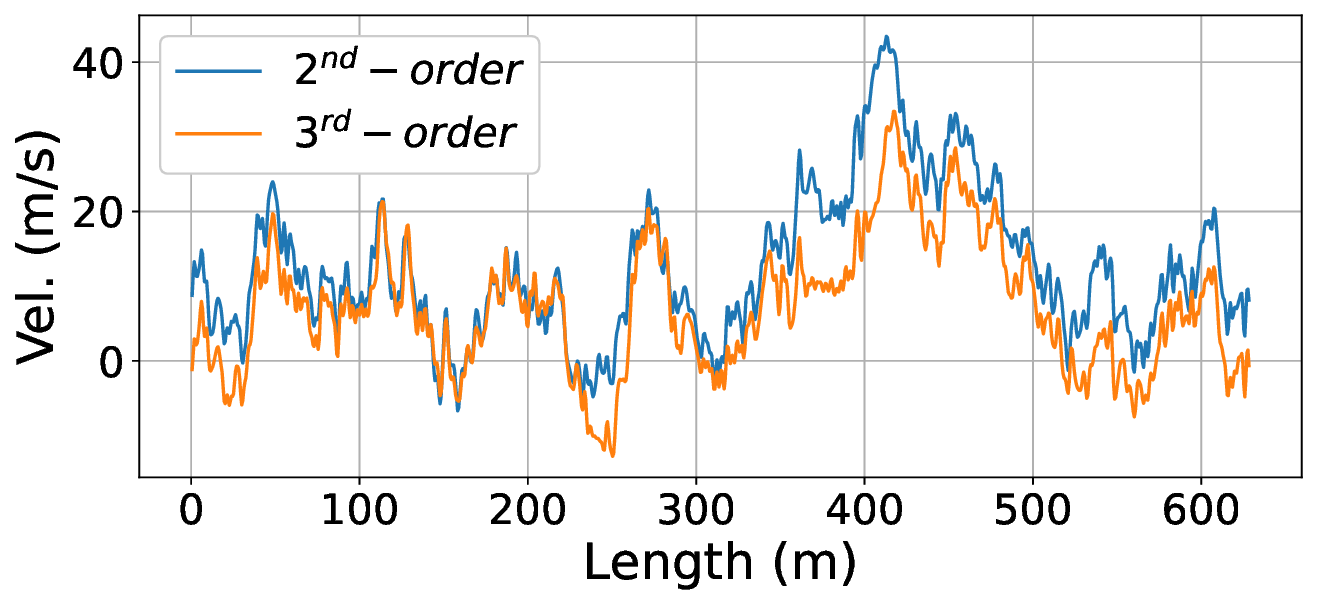}
        \caption{$t=127$ sec.}
        \end{subfigure}
        \begin{subfigure}{\textwidth}
        \includegraphics[width=1.0\linewidth]{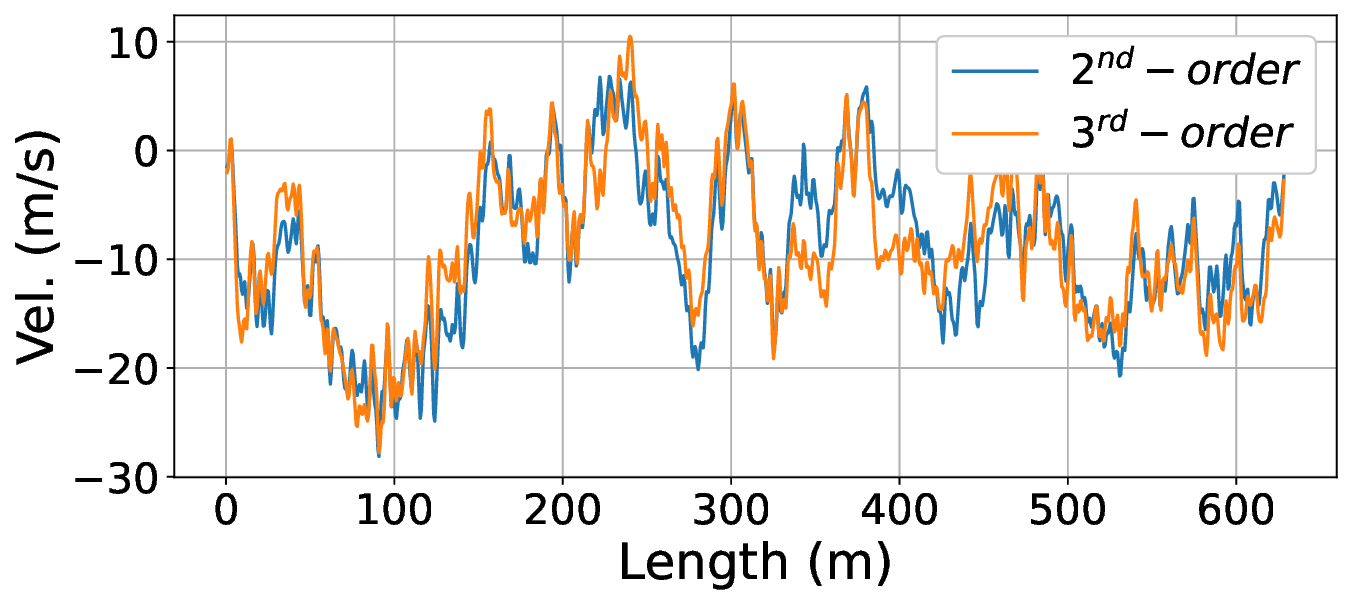}
        \caption{$t=254$ sec.}
        \end{subfigure}
    \end{subfigure}
    \caption{(a)--(c) Wind velocity time histories at three locations along the length of the bridge; (d)--(f) Wind velocity field along the length of the bridge at three time instances.}
    \label{fig:slice_along_time_and_length}
\end{figure}

We then simulate the stochastic wave in space-time using both the $2^{nd}$-and $3^{rd}$-order SRM. A sample realization of the entire stochastic wave simulated by the $2^{nd}$ and $3^{rd}$-order SRM is shown in Figure \ref{fig:sample_plots}. These samples are generated having identical random phase angles, which allows for a one-to-one comparison between the methods. The difference between the samples is shown for perspective in Figure \ref{fig:sample_plots}c. Here, we can see visible differences between the $2^{nd}$-and $3^{rd}$-order realizations, which are particularly noticeable in the peaks.
Figure \ref{fig:slice_along_time_and_length} further shows the wind velocity time history at different points along the length of the bridge and the velocity profile along the span of the bridge at different instances of time for the samples generated by $2^{nd}$ and $3^{rd}$ order Spectral Representation Method.
Again, we can see clear differences between these realizations, particularly in the peaks.

\begin{figure}[!ht]
\centering
\begin{subfigure}{.49\textwidth}
  \centering
  \includegraphics[width=1.0\linewidth]{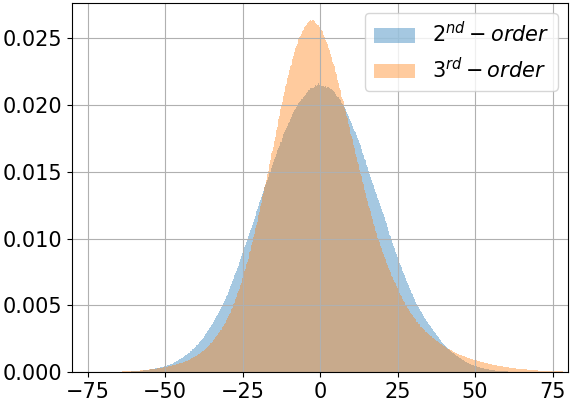}
  \label{fig:sample_pdf}
  \caption{}
 \end{subfigure}
 \begin{subfigure}{.46\textwidth}
  \centering
  \includegraphics[width=1.0\linewidth]{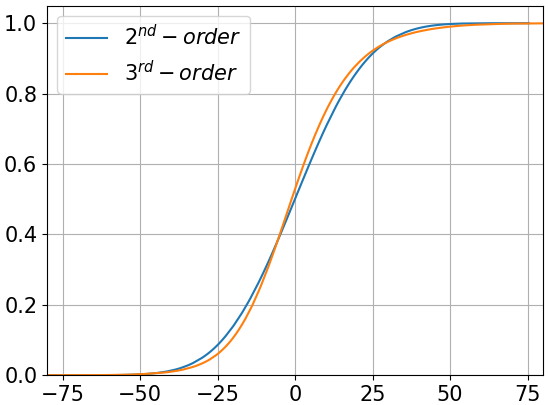}
  \label{fig:sample_cdf}
  \caption{}
 \end{subfigure}
 \caption{Empirical (a) PDFs and (b) CDFs from 1000 sample realizations of the wind field stochastic wave showing that the $2^{nd}$-order SRM clearly produces Gaussian samples while the $3^{rd}$-order SRM produces positively skewed sample functions.}
\label{fig:samples_distributions}
\end{figure}

Next, we simulated 1000 realizations of the $2^{nd}$ and $3^{rd}$-order stochastic waves to estimate their Probability Density Function (PDF), Cumulative Density Function (CDF), and sample statistics. The sample PDFs and CDFS for both the Gaussian and non-Gaussian waves are plotted in Figure \ref{fig:samples_distributions}. We can clearly see that the PDF of the non-Gaussian ($3^{rd}$-order) wave is positively skewed, having a peak that is tilted slightly to the left and a heavier upper tail than the Gaussian ($2^{nd}$-order) wave.

\captionsetup[subfigure]{font=small,skip=-15pt}
\captionsetup[figure]{font=small,skip=0pt}

This is verified by the sample statistics shown in Table \ref{tab:statistics_table} where we can see that both the samples generated by $2^{nd}$ and $3^{rd}$-order SRM can match variance accurately, only the $3^{rd}$-order Spectral Representation Method can capture the skewness of the process.
\begin{table}[!ht]
    \centering
    \begin{tabular}{|c|c|c|c|}
        \hline
        & \textbf{$2^{nd}$-order} & \textbf{$3^{rd}$-order} & \textbf{Theoretical} \\
        \hline
        Mean &  0.00 & 0.00 & 0.00 \\
        Variance & 336.23 & 337.64 & 340.41 \\
        Skewness & 0.016 & 0.442 & 0.497 \\
         \hline
    \end{tabular}
    \caption{Statistics of the simulated waves by the $2^{nd}$ and $3^{rd}$ Spectral Representation Method}
    \label{tab:statistics_table}
\end{table}
We further note that a true validation of the $3^{rd}$-order SRM requires a comparison of the sample bispectrum and the target bispectrum from Eq.\ \eqref{eqn:bispectrum}. Because the bispectrum is 4-dimensional, it cannot be visually compared here. However, prior works have demonstrated that the $3^{rd}$-order SRM accurately matches the bispectrum (both theoretically and in its sample properties) \cite{Shields2017}, and these properties translate directly for this extension. This is verified by the fact that the sample realizations match the target skewness, which is derived by integrating the bispectrum.


\section{Conclusion}

A methodology is proposed to simulate a non-Gaussian wind velocity field along the length of long-span structures using $3^{rd}$-order stochastic waves. By modeling the wind field as a stochastic wave, the methodology avoids the numerical \textcolor{blue}{, memory} and computational pitfalls of modeling stochastic vector processes with many (hundreds to thousands) of components. By modeling the wind-field using a novel extension of the $3^{rd}$-order Spectral Representation Method, the methodology is capable of capturing the skewed non-Gaussian character of the wind field. We further propose a computationally efficient implementation of the proposed $3^{rd}$-order SRM for stochastic waves that leverages the fast Fourier transform.

The methodology is then applied for simulation of the wind velocity field along the span of a long-span bridge having length 628 meters with spatial resolution of 0.6 meters and temporal resolution of 0.12 sec. Simulations of the stochastic wave demonstrate that it matches the $2^{nd}$ and $3^{rd}$-order properties of the target wind field with high accuracy and comparison with $2^{nd}$-order simulations shows the contrast between the proposed approach and a Gaussian wind field. The wind field simulated here has an arbitrarily prescribed analytical bispectrum, but recent advances in field data collection and wind tunnel modeling \cite{catarelli2020automation} suggest that wind bispectra could be obtained from real wind velocity data, as they have e.g.\ for ocean wave data \cite{elgar1985observations}, and integrated into wind field simulations using the proposed method. 

\section{Acknowledgements}

This material is based upon work supported by the National Science Foundation under Grant No. 1652044.

\bibliographystyle{model1-num-names}
\bibliography{elsarticle-template-1-num.bib}

\end{document}